\documentclass[11pt,a4paper]{llncs}
\usepackage{amsmath}
\usepackage{amssymb}
\usepackage{graphicx}
\usepackage{marvosym}
\usepackage{url}
\usepackage{fancyhdr}

\usepackage{geometry}
\newcommand{\keywords}[1]{\par\addvspace\baselineskip
\noindent\keywordname\enspace\ignorespaces#1}

\fancypagestyle{firstpage}{\fancyhf{}
}

\begin{document}


\title{\LARGE{Integrating Large Language Models in Financial Investments and Market Analysis: A Survey}}


%
%
\author{\large{Sedigheh Mahdavi  \and Jiating (Kristin) Chen \and Pradeep Kumar Joshi \and Lina Huertas Guativa\and Upmanyu Singh}}
\institute{\large{AI Research Lab, Blend360,\\ Columbia, USA}}

\maketitle

\thispagestyle{firstpage}

\begin{abstract}
Large Language Models (LLMs) have been employed in financial decision making,  enhancing analytical capabilities for investment strategies. Traditional investment strategies often utilize quantitative models, fundamental analysis, and technical indicators.
However, LLMs have introduced new capabilities to process and analyze large volumes of structured and unstructured data, extract meaningful
insights, and enhance decision-making in real-time. This survey provides a structured overview of recent research on LLMs within the financial domain, categorizing research contributions into four main frameworks: LLM-based Frameworks and Pipelines, Hybrid Integration Methods, Fine-Tuning and Adaptation Approaches, and Agent-Based Architectures. This study provides a structured review of recent LLMs research on applications in stock selection, risk assessment, sentiment analysis, trading, and financial forecasting. By reviewing the existing literature, this study highlights the capabilities, challenges, and potential directions of LLMs in financial markets.
\end{abstract}

\section{Introduction}

Large Language Models (LLMs) have been utilized in financial decision making, such as stock selection, risk assessment, sentiment analysis, trading, and financial forecasting.
Traditional financial analysis generally struggles to incorporate the unstructured information contained in news articles, social media, corporate filings, and analyst reports that significantly influence market movements. LLMs present a potential solution to this challenge by enabling the integration of textual and numerical data for a more comprehensive financial analysis. In this survey, we classified the current approaches into four main frameworks: (1) LLM-based Frameworks and Pipelines, which implement systematic architectures for financial analysis; (2) Hybrid Integration Methods, which combine traditional financial models with LLM capabilities; (3) Fine-Tuning and Adaptation Approaches, which customize LLMs for specific financial tasks; and (4) Agent-Based Architectures, which utilize multiple AI agents for collaborative decision making.   

Recent research has utilized several key techniques, such as Retrieval-Augmented Generation (RAG), which enhances LLM outputs by incorporating proprietary financial data, addressing the limitation of outdated knowledge in pre-trained models; Chain-of-Thought (CoT) reasoning, which improves transparency in financial decision-making by breaking down complex analyses into sequential steps; In-Context Learning (ICL), which enables models to adapt to novel financial tasks without parameter updates; and Parameter-Efficient Fine-Tuning (PEFT) and Low-Rank Adaptation (LoRA), which provide resource-efficient methods for customizing LLMs to specific financial domains. The datasets utilized in these reviewed studies include a wide range of datasets from established market indices, such as the s\&p 500, to specialized financial news collections. These diverse data sources allow researchers to evaluate LLM performance across different market conditions, sectors, and periods, providing insights into the robustness of LLM-based investment strategies.  

The remainder of the paper is structured as follows. Section 2 presents the main techniques that have been utilized to enhance the performance of LLMs. Section 3 reviews LLMs-based research for financial decision-making systems. Section 4 concludes the paper with findings and future directions. 

\section{Background Review}
This section provides an overview of the main approaches that have been widely developed in recent studies. 

\subsection{Retrieval-Augmented Generation }
Retrieval-Augmented Generation (RAG) \cite{Lewis2020} is a technique that enhances the performance of generative AI models by integrating the most current and relevant proprietary data with an existing LLM. The RAG retrieves context from both structured and unstructured data sources to provide accurate and domain-specific information for response generation. 

\subsection{Chain-of-Thought}
Chain-of-Thought (CoT) reasoning \cite{Wei2022} is a structured approach to enhance the interpretability and decision-making abilities of LLMs by guiding them to think step-by-step through a task. Rather than providing direct answers, CoT reasoning encourages models to break down problems into smaller sequential steps, mimicking human analytical reasoning. 

\subsection{In-Context Learning}
In-Context Learning (ICL) \cite{Brown2020} is a capability of LLMs that enables them to generalize to novel tasks without explicit parameter updates. ICL facilitates a model's learning from examples provided within the input prompt. By conditioning on a limited number of demonstrations, typically formatted as input-output pairs, the model infers task patterns and generates appropriate responses. 

\subsection{Low-Rank Adaptation}

Fine-tuning with Low-Rank Adaptation (LoRA) \cite{Hu2022} is a parameter-efficient method for adapting LLMs to specific tasks or domains without retraining the entire model. Instead of updating all model parameters during training, LoRA introduces low-rank trainable matrices into the existing layers of the model. These matrices learn task-specific knowledge while keeping the original model weights frozen, making the process computationally efficient and memory friendly. 

\subsection{Parameter-Efficient Fine-Tuning }
Parameter-Efficient Fine-Tuning (PEFT) \cite{Houlsby2019,Li2021} is an advanced technique designed to fine-tune LLMs by updating only a small subset of their parameters, thereby significantly reducing computational costs, memory usage, and training time. Several methods are commonly employed in PEFT:  adapters (small task-specific modules inserted between model layers), LoRA, prefix-tuning (learnable input embeddings for behavior modification), and bitFit (bias-term adjustment for performance optimization).  
\subsection{Reinforcement Learning with Human Feedback }
Reinforcement Learning with Human Feedback (RLHF) \cite{Radford2021,Christiano2017} is an advanced technique that combines traditional reinforcement learning (RL) with human-generated feedback to guide the model training process. RLHF utilizes human input, qualitative or quantitative evaluations of the model's outputs, as a feedback signal. This human-provided feedback helps the model adjust its behavior, aligning it more closely with human preferences or specific task requirements. The RLHF process generally involves three stages: 1- Supervised Fine-Tuning (SFT): The model is first trained on a dataset of labeled examples, providing a foundational understanding before further refinement. 2- Reward Model Training: In this stage, a reward model is trained using human feedback to evaluate the outputs generated by the model. Humans rank multiple responses, and these rankings are used to train the reward model, helping it predict user preferences more effectively. 3- Policy Optimization: The model refined through using reinforcement learning. In this stage, the reward model provides continuous feedback to help iteratively adjust the model's policy.  
\subsection{Mixture of Experts}
The mixture-of-experts (MoE) framework \cite{Jacobs1991} was introduced to improve scalability and enable multitask learning, particularly in areas such as computer vision and natural language processing. Using a combination of specialized models, the MoE allows for more efficient computation and improved performance by dynamically routing input to the most appropriate experts based on the task at hand. 

\section{Materials and Methods}
This section provides a brief overview of research efforts focused on integrating LLMs into the financial domain. These approaches are categorized into four main categories: LLM-based Frameworks and Pipelines, which focus on the systematic implementation of LLMs through either modular architectures or sequential processing steps; Hybrid Integration Methods, which combine traditional financial analysis with AI capabilities; Fine-Tuning and Adaptation Approaches, which customize LLMs for specific financial tasks; and Agent-Based Architectures, which leverage multiple AI agents for complex decision making. The following subsections provide a brief overview of LLM–based techniques, and Tables 1-4 summarize the model, data, and application for the four main categories.


\subsection{LLM-based Frameworks and Pipelines }
\begin{table}[htbp]
\caption{LLM-based Frameworks and Pipelines: Models, and Data, Applications}
\begin{center}
\begin{tabular}{|p{0.2\textwidth}|p{0.26\textwidth}|p{0.3\textwidth}|p{0.3\textwidth}|}
\hline
\textbf{Method/Paper} & \textbf{LLM Model} & \textbf{Data} & \textbf{Application} \\
\hline
MarketSenseAI \cite{Fatouros2024} & GPT-4 & S\&P 100 stocks & Stock selection and investment insight \\
\hline
Ploutos \cite{Tong2024} & GPT-4 & ACL18 and CIKM18 & Stock movement \\
\hline
ChainBuddy \cite{Zhang2024a} & GPT-based models & \raggedright Common Crawl, Wikipedia, and domain-specific corpora & Financial market analysis \\
\hline
GPT-InvestAR \cite{Gupta2023} & GPT3.5 & S\&P 500 index & Stock investment \\
\hline
LLMoE \cite{Liu2025} & Llama3.2 & MSFT-AAPL data & Trading strategies \\
\hline
\cite{Lee2024a} & GPT-4-Turbo, LLaMA3 & KOSPI200 index & Stock market prediction \\
\hline
\end{tabular}
\end{center}
\end{table}

In \cite{Fatouros2024}, MarketSenseAI, an innovative AI-driven framework utilized GPT-4’s advanced reasoning capabilities to revolutionize stock selection and portfolio management in financial markets. By integrating CoT and ICL techniques, MarketSenseAI synthesizes insights from diverse data sources, including market trends, news, financial fundamentals, and macroeconomic factors as an expert investment decision-making. The framework consists of modular components, including a news summarizer, financial fundamentals analyzer, stock price dynamics module, and macroeconomic environment summarizer. These components work together to analyze company-specific news, evaluate recent financial reports, compare stock price trends with peers, and assess global economic conditions. The outputs are integrated into a final decision-making layer, where GPT-4 acts as an expert analyst to produce actionable investment signals (e.g., "buy," "sell," or "hold") accompanied by detailed explanations. The prompt structure encompasses news summaries, price dynamics, the macroeconomic environment, and the fundamental context, enabling GPT-4 to analyze each stock's specific context while applying its general financial knowledge. Empirical evaluation of s\&p 100 stocks over a 15-month period demonstrates the performance of framework, achieving up to $72\%$ cumulative returns and delivering an excess alpha between 10\% and 30\%. 

A novel financial LLM framework (Ploutos) was introduced in \cite{Tong2024} for interpretable stock movement prediction. Ploutos addresses two key challenges: (1) integrating both textual and numerical data to enhance predictive accuracy, and (2) providing clear, interpretable rationales for stock movements. The framework consists of two core components, PloutosGen and PloutosGPT. PloutosGen processes multimodal financial data by incorporating textual sources (e.g., news articles and tweets) and numerical information (e.g., stock prices and financial ratios). It utilizes a pool of specialized experts, including a technical analysis expert, which integrates multi-feature time-series data with textual explanations using a number-to-text alignment mechanism, and human experts, which encode financial analysts' insights into macroeconomic indicators and market sentiments. PloutosGPT synthesizes insights from PloutosGen and generates interpretable reasons for stock-movement predictions. It employs two novel techniques: rearview-mirror prompting, which enhances interpretability by contextualizing predictions based on historical trends, and dynamic token weighting, which improves the prediction clarity by emphasizing key tokens based on cosine similarity with learned embeddings. The framework was evaluated on two public datasets, ACL18 and Twitter stock dataset, for stock movement prediction. 

Jingyue et al. \cite{Zhang2024a} introduced ChainBuddy, an AI-driven system, that automates the creation and management of LLM workflow. Addressing the "blank page problem" enables users to generate interactive and customized pipelines using natural language inputs, thereby reducing technical barriers for non-experts. The system operates in three stages: input, processing, and output. In the input stage, users provide a natural language description of their desired workflow. In the processing stage, the AI agent applies reasoning algorithms to infer the user intent and map it to relevant LLM components. In the output stage, a fully configured pipeline is generated that allows users to execute or refine it iteratively. ChainBuddy’s approach modularized LLM functionalities as standalone components that are dynamically combined and connected via standardized interfaces, ensuring efficient workflow execution. ChainBuddy relies on LLM training datasets, user-interaction data, and modular components to generate a workflow. Pre-trained LLMs trained on corpora such as Common Crawl and Wikipedia support various NLP tasks. User queries and feedback refine pipeline configurations dynamically, while predefined LLM PyTorch and TensorFlow tools serve as reusable building blocks for workflow assembly. The system is applied to automated LLM management, workflow customization, and rapid AI prototyping, allowing users to create workflows for tasks such as summarization, question-answering, and financial analysis. 

GPT-InvestAR was introduced in \cite{Gupta2023}, a framework that utilizes LLMs to extract insights from corporate annual reports ($10-K$ filings) and improve stock investment strategies. The method involves five key steps: (1) Accessing Annual Reports: Historical $10-K$ filings for the top 1500 companies by market capitalization are collected from the SEC archives. (2) Generating Document Embeddings: Relevant sections of the reports are identified using cosine similarity based on vector embeddings and stored in ChromaDB for efficient querying. (3) Feature Generation with LLM: GPT-3.5-Turbo was used to generate financial features by asking specific questions about company health, such as growth strategies. Responses and confidence scores were used as model features. (4) Label Creation: Stock returns (percentage change between filing dates) are calculated for each company and the s\&p 500 index, generating labels for stock performance. (5) Training the Machine Learning Model: A regression model trained on data from 2002 to 2017 and tested on data from 2018 to 2023 uses LLM-generated features to predict future stock performance. This approach outperforms the s\&p 500 index, demonstrating the potential of combining LLM-driven insights with machine learning to enhance investment decision making. Liu et al., \cite{Liu2025} presented LLMoE, a novel framework for stock market prediction that integrates LLMs as dynamic routers within an MoE architecture. Traditional financial models often face challenges due to static routing mechanisms and lack of textual data. LLMoE addresses these limitations by incorporating both numerical stock data and textual news, allowing for more effective expert selection and improved market predictions. The framework consists of three main stages: a) an LLM-based router that processes multimodal inputs and selects appropriate experts, b) expert models that make predictions for different market conditions, and c) a trading strategy based on expert outputs using an "All-in All-out" approach. This strategy involves fully investing available funds when positive price movements are predicted and selling all holdings when negative movements are predicted, aiming to maximize returns based on expert predictions. Experiments conducted on MSFT and AAPL datasets spanning 2006 to 2016 demonstrated LLMoE's superior performance compared to baseline models and achieved significant improvements of over 25\% in key return metrics.

Lee et al. \cite{Lee2024a} proposed a novel framework that used LLM to transform qualitative financial disclosures into structured signals for stock market prediction. This method uses daily investor-facing reports and constructs a pipeline that maps free-text insights into the numerical factors that can be used for downstream predictive tasks. The framework consists of six main stages: (1) Data collection from Naver Finance ensures selection of the most relevant investor-focused reports; (2) Multiple trial generation introduces low-temperature sampling to stabilize LLM outputs; (3) Factor generation uses Finance-LLaMA-8B to extract financial factors; (4) Autoregressive moving shot combines these factors with historical stock data for few-shot inference; (5) Factor scoring prompts apply a Likert-style scale to produce sentiment scores; and (6) Value-to-score scaling ensures alignment with real financial distributions. This pipeline effectively integrates unstructured qualitative content with numerical modeling, without requiring external textual corpora or labeled training data. The framework was evaluated using proprietary financial reports and KOSPI200 market data, which include daily closing prices of the top 200 companies listed on the Korean Exchange. The results show that this approach achieves higher predictive accuracy and improved robustness compared to baseline models, particularly in contexts where traditional numeric-only forecasting methods may underperform.      

\begin{table}[htbp]
\caption{Hybrid Integration Methods: Models, and Data, Applications}
\begin{center}
\begin{tabular}{|p{0.24\textwidth}|p{0.25\textwidth}|p{0.3\textwidth}|p{0.3\textwidth}|}
\hline
\textbf{Method/Paper} & \textbf{LLM Model} & \textbf{Data} & \textbf{Application} \\
\hline
ChatGPT-based Investment Portfolio Selection \cite{Romanko2023} & ChatGPT & S\&P500 market & Investment portfolio selection \\
\hline
LLM-based stock market trend prediction \cite{Swamy2023} & ChatGPT & \raggedright Historical stock trends, news, and trading-volumes sentiment data & Stock market trend \\
\hline
FLLM \cite{Chu2023} & GPT-3, GPT-4, Llama & Proprietary dataset & Financial analysis \\
\hline
Few-Shot Stock Trend Prediction \cite{Deng2024} & ChatGPT and GPT-3 & tS\&P 500 index and CSI-100 index & Stock trends \\
\hline
LLM-Augmented Linear Transformer--CNN \cite{Zhou2025} & \raggedright ChatGPT-4o, FinBERT & S\&P 500 and financial reports & Stock price prediction \\
\hline
MuSA \cite{Li2025a} & FinBert & DJIA and S\&P 500 & Financial Portfolio Optimisation \\
\hline
SEP \cite{Koa2024} & \raggedright OpenAI GPT-3.5-turbo-16k and Vicuna-13b-v1.5-16k & ACL18 StockNet & Stock predictions \\
\hline
\cite{Joshi2025} & \raggedright ChatGPT-4 and Google Gemini & Credit risk and market risk assessment data & Financial market integrity and risk management \\
\hline
Hybrid LLM-based framework \cite{Elahi2024} & GPT-3, GPT-4, LLaMA-2, LLaMA-3 & Financial reports (10-K) & Stock price movement prediction \\
\hline
\cite{Papasotiriou2024} & GPT-4-32k & S\&P 500 & Stock rating predictions \\
\hline
\end{tabular}
\end{center}
\end{table}

\subsection{Hybrid Integration Methods }
Romanko et al. \cite{Romanko2023} developed the ChatGPT-based Investment Portfolio Selection methodology, a novel approach to investment portfolio construction by integrating ChatGPT stock selection capabilities with traditional quantitative portfolio optimization methods. Initially, ChatGPT was used to select a universe of stocks from the s\&p 500, with a process involving 30 requests to the model. Using a majority voting approach, a reliable stock universe was built based on ChatGPT selection. The selected stocks were then incorporated into the portfolio construction using two strategies: 1) an equally weighted portfolio, where all selected stocks were assigned equal weights, and 2) a GPT-weighted portfolio, where weights were determined by another GPT-4 prompt based on ChatGPT stock selection rationale. In addition, classical portfolio optimization techniques were applied to construct three optimized portfolios: the minimum variance portfolio, which seeks to minimize overall risk; the maximum return portfolio, which aims to maximize expected returns; and the maximum Sharpe ratio portfolio, which balances risk and returns to achieve the best risk-adjusted returns. The problem is then formulated as a mixed-integer optimization task using binary variables with cardinality restrictions to ensure that a fixed number of assets are included. They further compared the out-of-sample performance of eight portfolios derived from three types of strategies, including both GPT-weighted and equally weighted portfolios as well as portfolios optimized using traditional techniques with and without cardinality constraints.

Swamy et al. \cite{Swamy2023} introduced a novel approach to predict stock market trend using LLMs to integrate investor sentiment with traditional quantitative analysis. Investor sentiment is captured through sentiment analysis of news articles, which is processed using APIs such as StockNewsAPI with LSTM-based sentiment classification. Each news item is assigned a sentiment score—positive, neutral, or negative— that focuses on news released before market opening and during trading hours. In addition to sentiment analysis, they employed moving averages (MAs) of stock prices over different periods (10, 50, and 200 days) to identify short-term fluctuations and long-term trends. Various combinations of MAs were tested, and the results indicated that a combination of MAs over different periods provided optimal predictive performance without overcomplicating the model. To further enhance the prediction accuracy, the options volume and Volatility Index (VIX) are incorporated. Options trading volume is a key measure of investor outlook, whereas the VIX reflects overall market sentiment. By combining sentiment analysis, moving averages, options trading volumes, and dependency factors, the proposed methodology constructs a robust dataset for stock trend predictions. LLMs process these aggregated data using prompt engineering techniques to map sentiments and market indicators to structured financial features. 

A data-centric approach \cite{Chu2023} was introduced to enhance the efficacy of LLMs for financial tasks by addressing their limitations in integrating and reasoning complex financial data. Instead of directly overloading LLMs with raw text, the financial large-language model (FLLM) utilizes multitask prompt-based fine-tuning to preprocess financial texts, ensuring enhanced contextual understanding. The methodology involves modular components for event matching, viewpoint evaluation, and keypoint extraction, enabling the FLLM to refine its financial insights before generating the final interpretations. A key innovation is abductive augmentation reasoning, which automatically generates high-quality training data by modifying pseudo-labels from the model’s outputs, thereby mitigating the scarcity of labeled financial datasets. For empirical evaluation, this study utilizes a proprietary dataset sourced from three main channels: web crawling (real-time financial news and market analysis), purchased industry reports, and in-house data from platform discussions and influencer perspectives. This dataset consists of millions of daily financial text entries preprocessed using domain-specific parsing techniques and event-correlation models. This study also introduces an open-source benchmark for financial text analysis, demonstrating the advantages of a data-centric AI approach in finance. The experimental results indicate that FLLM outperforms traditional financial LLMs, achieving state-of-the-art performance in financial analysis, interpretation, and reasoning tasks, highlighting the effectiveness of data-driven optimization in financial AI applications.

In \cite{Deng2024}, an enhanced few-shot setting was used to enhance stock trend predictions using large-language models. They proposed a two-step "denoising-then-voting" approach to reduce the need for large amounts of labeled data. First, the model classifies individual news articles as "Up," "Down," or "Irrelevant" using a few-shot prompt. This step filters out irrelevant news and reduces noise. Second, predictions from relevant news are aggregated using majority voting to determine the final stock trend. By handling news individually, rather than merging it, this method prevents information loss and ensures a more reliable prediction process. The proposed method was tested on financial news datasets related to the s\&p 500 index, the CSI-100 index (A-share stocks from China), and Hong Kong stocks (companies listed on the Hong Kong Stock Exchange). Zhou et al. \cite{Zhou2025} introduces a hybrid deep learning framework that integrates an LLM, a Linear Transformer, and a CNN to enhance stock price prediction by integrating temporal, spatial, and contextual features. It leverages LLM-generated financial insights, visual patterns from stock charts, and numerical indicators to improve the predictive accuracy beyond traditional forecasting methods. The model is designed for stock price prediction, combining structured AI-based trading signals with multimodal data sources to better capture the market dynamics. The framework consists of a Linear Transformer for processing historical stock prices, a Stock Chart-CNN (SC-CNN) for extracting spatial patterns from candlestick charts, and an LLM branch where ChatGPT-4o generates daily financial summaries that are transformed into numerical vectors using FinBERT. These features were concatenated and processed through fully connected layers for stock prediction. This study utilizes s\&p 500 stock data from 2022 to 2023, incorporating historical price data for time-series analysis, candlestick chart images for CNN-based pattern recognition, and technical indicators enriched by LLM-generated reports. This structured dataset enabled the model to process diverse financial signals and generate predictions with improved interpretability. Performance evaluation was conducted using the Root Mean Squared Error (RMSE), Mean Absolute Percentage Error (MAPE), and Mean Absolute Error (MAE), and the results demonstrated that the proposed model significantly outperformed traditional and hybrid baselines. 

In \cite{Li2025a}, MuSA (Multimodal and Sentiment-Based Adaptive), an advanced AI-driven trading framework, was introduced to optimize financial portfolio management by integrating LLMs, deep reinforcement learning (DRL), and sentiment analysis. By leveraging FinBERT for sentiment extraction and an entropy-based confidence learning mechanism to filter unreliable news sources, MuSA enhances the reliability of market-sentiment signals. Additionally, the framework incorporates Directional Change (DC) and twin-delayed deep deterministic policy gradient (TD3) to predict price dynamics using both time-series and event-based perspectives. MuSA consists of modular components, including a sentiment analyzer, price analyzer, and portfolio aggregator, which work together to extract and validate market sentiment, predict future price trends, and generate optimized portfolio weights that dynamically adapt to market volatility. These outputs are integrated into a decision-making layer where DRL-based optimization techniques produce portfolio allocations that maximize returns while managing risk. The empirical evaluation of DJIA (February 2010 to May 2020) and s\&p 500 indices (2010-2020) demonstrates MuSA’s superior performance, achieving higher annualized returns and Sharpe ratios compared to algorithmic and DRL-based baselines.

In \cite{Koa2024}, the Summarize-Explain-Predict (SEP) framework was introduced as an innovative approach to explainable stock prediction by integrating self-reflective learning and RL with LLMs. Traditional deep learning models and LLMs struggle with chaotic social text data and lack annotated explanations, making stock predictions less interpretable. SEP addresses these limitations by leveraging a self-reflective agent that enables LLMs to autonomously refine their predictions and explanations without requiring expensive human-labeled data. The framework consists of three key modules: Summarize, Explain, and Predict. The Summarize module extracts concise factual summaries from unstructured social text data, the Explain module generates stock movement predictions with interpretable explanations using a self-correcting process, and the Predict module fine-tunes the model using reinforcement learning with Proximal Policy Optimization (PPO) to maximize prediction accuracy. The framework was trained and evaluated using an updated version of the ACL18 StockNet dataset (2020–2022), which includes price data from Yahoo Finance and tweets collected via the Twitter API, ensuring a comprehensive foundation for learning text-driven stock market signals. Empirical evaluation demonstrates that SEP outperforms traditional deep learning models and financial LLM baselines in terms of both predictive accuracy and explanation quality. 
 
In \cite{Joshi2025}, the role of prompt engineering was investigated to optimize the effectiveness of LLMs such as ChatGPT-4 and Google Gemini for financial market integrity and risk management. This study first provides a detailed literature review on the increasing use of prompt engineering in the financial sector. This study evaluates the impact of different prompt configurations on financial decision making. Researchers generated 50 questions designed as queries for backend financial models and then assessed both the accuracy of the responses and the number of prompts required to obtain reliable results. To ensure regulatory compliance, the LLMs were instructed to source information exclusively from .gov websites. Additionally, a panel of three financial analysts reviewed the questions to verify their computational relevance and alignment with the financial models. This methodology was applied to both credit risk and market risk assessments to measure how prompt engineering enhances AI model performance. 

A hybrid LLM-based framework \cite{Elahi2024} was proposed to improve stock price movement prediction by combining structured financial data with unstructured textual information. Utilizing the contextual understanding of LLMs, such as GPT-3, GPT-4, LLaMA-2, and LLaMA-3, enables reasoning across both numerical and narrative inputs. The method first retrieves relevant news articles for each target company by using RAG. These news articles, along with financial indicators, such as revenue, net income, free cash flow, and total assets, are combined into structured prompts. These prompts are then fed into the LLM in zero-shot, two-shot, and four-shot settings, enabling the model to learn from contextual examples and produce predictions about the stock’s future movement, classified as either “UP” or “DOWN,” over 3- and 6-month intervals. The framework was evaluated on a proprietary dataset consisting of 20 highly traded U.S. companies, incorporating approximately 5,000 financial news articles collected from October 2021 to January 2024. The results showed that GPT-4 outperformed all other models, achieving the highest accuracy and consistency in predictions. 

Kassiani Papasotiriou, et. al. \cite{Papasotiriou2024} used LLMs to predict stock performance and generate stock ratings. GPT-4-32k (v0613) was selected, and then the model was instructed to act as a financial analyst, providing role clarity, and was given financial context by defining stock rating scales, synonyms, and financial fundamentals to ensure consistency. CoT and few-shot prompting were employed to enhance reasoning, and input data were first structured with textual information, followed by numerical tables for optimized comprehension. A Chain of Verification (CoVE) was used to ensure that predictions aligned with the correct dates. The authors framed stock rating prediction as an ordinal classification task, aiming to classify stocks as Strong Sell, Moderate Sell, Hold, Moderate Buy, or Strong Buy for future time horizons of 1, 3, 6, 12, and 18 months. The datasets included stock ratings for s\&p 500 companies, news articles filtered using named entity recognition (NER) and summarized by GPT-4-32k to highlight key events and trends, sentiment analysis of these summaries, historical stock prices, quarterly financial fundamentals detailing balance sheets, income statements, and cash flow. The model’s performance was evaluated by comparing predicted ratings with actual stock performance over these horizons. The performance was determined by dividing companies’ forward returns into quintiles, with the bottom quintile corresponding to a Strong Sell and the top quintile to a Strong Buy. Performance was assessed using the MAE between predicted and actual ratings.

\begin{table}[!ht]
\caption{Fine-Tuning and Adaptation Approaches: Models, and Data, Applications}
\begin{center}
\begin{tabular}{|p{0.23\textwidth}|p{0.22\textwidth}|p{0.35\textwidth}|p{0.3\textwidth}|}
\hline
\textbf{Method/Paper} & \textbf{LLM Model} & \textbf{Data} & \textbf{Application} \\
\hline
Multimodal Gen-AI \cite{Li2023} & Llama2 and GPT-3.5 & Goldman Sachs reports, stock news data, Benzinga stock price, and Yfinance data & AI-assisted investment analysis \\
\hline
Fine-tuning LLMs \cite{Guo2024} & \raggedright DeBERTa, Mistral, Llama & Financial newsflow & Stock return prediction \\
\hline
Finance-specific LLMs \cite{Chiu2025} & LLaMA-2 7B & Stanford Alpaca, Financial PhraseBank, Twitter Financial News Sentiment, Climate Sentiment, and MD\&A  & \begin{minipage}{0.3\textwidth}Financial sentiment analysis and return prediction\end{minipage} \\
\hline
Distilled LLM based text-classification \cite{Bhat2024} & Distilled RoBERT & Financial news data & Stock price trends \\
\hline
\cite{Ni2024} & llama-3-8b-Instruct-4bit, GPT-4 & S\&P 500, financial reports, analyst sentiment, and market indices & Predicting stock movements \\
\hline
Stock-Chain framework \cite{Li2024} & StockGPT & AlphaFin & Financial analysis \\
\hline
FinLlama \cite{Iacovides2024} & Llama 2 7B & \raggedright Financial news and Earnings reports & Trading strategies \\
\hline
StockTime \cite{Wang2024b} & LLaMA3-8B & \raggedright S\&P- Bigdata23, Bigdata22, ACL18, and CIKM1 & Stock price prediction \\
\hline
SAPPO \cite{Kirtac2025} & LLaMA 3.3 & Refinitiv financial news & Portfolio optimization \\
\hline
\end{tabular}
\end{center}
\end{table}

\subsection{Fine-Tuning and Adaptation Approaches}
Li et al. \cite{Li2023} proposed fine-tuning Llama2 and GPT-3.5 to automate financial investment research using domain-specific datasets, including research reports, market news, and time-series data. The study utilized unsupervised, supervised, and instruction-based LoRA fine-tuning methodologies: Goldman Sachs reports were employed for unsupervised fine-tuning, while supervised fine-tuning utilized stock news data obtained from Benzinga (available on Kaggle) and stock price data retrieved via the Yfinance API to examine the impact of news on stock prices. The results indicate that fine-tuning significantly enhanced Llama2's capacity to process investment-related queries and generate insights, although both the baseline and fine-tuned versions struggled to handle complex or novel queries. By contrast, instruction fine-tuned GPT-3.5 surpassed its baseline by providing clearer, more direct, and logically consistent answers, aligning more closely with human analysts' expectations. This fine-tuning resulted in significant improvements in relevance, decision-making, and reasoning, making GPT-3.5 a more suitable candidate for AI-assisted investment analysis. 

\cite{Guo2024} utilized fine-tuning LLMs for stock return forecasting using financial news. They developed a forecasting framework consisting of two key components: (1) a text representation component that converts financial news into numerical embeddings using LLMs, and (2) a forecasting component that maps these embeddings to stock return predictions. They compared encoder-only (DeBERTa) and decoder-only (Mistral, Llama) LLMs to evaluate how their different text representations impact prediction performance. To integrate token-level representations into the forecasting component, two methods are explored: bottleneck representations, where the model compresses the entire news sequence into a single vector using an End-of-Sequence (EOS) token, and aggregated representations, where token embeddings are averaged to form the final representation. The models were fine-tuned using LoRA to optimize the computational efficiency. The experiments were conducted on company-level financial newsflow data from 2003 to 2019, with investment universe datasets from North American, European, and Emerging markets.  

Chiu et al. \cite{Chiu2025} developed a fine-tuned LLaMA-2 model with an AI-driven summarization process for enhanced financial sentiment analysis and return prediction. First, LLaMA-2 7B is fine-tuned on specialized financial datasets (Stanford Alpaca, Financial PhraseBank, Twitter Financial News Sentiment, Climate Sentiment). Then, to handle lengthy financial documents, a two-step AI-driven summarization process using Google’s LongT5 model employs iterative summarization with overlapping windows followed by aggregative summarization. The fine-tuned LLaMA-2 model analyzes the summarized texts using a three-category sentiment classification system with Bayesian framework. These sentiment signals are utilized to determine two return prediction strategies: (1) a trading strategy that buys stocks following positive sentiment and sells stocks following negative sentiment on day t+1, and (2) analysis of cumulative abnormal returns around filing dates using windows from [0,1] to [0,5] days. The performance of the fine-tuned LLaMA-2 model is compared against other models, including FinBERT and traditional machine learning approaches, using stock data from 1994 to 2022 sourced from CRSP and Compustat, along with MD\&A texts extracted via EDGAR in R. The results show that trading strategies based on LLaMA-2 sentiments produce significantly higher buy-and-hold returns  than other methods. 

In \cite{Bhat2024}, a Distilled LLM was fine-tuned to extract the emotional tone and intensity of financial news headlines to predict stock price trends. Financial news data were collected through APIs rather than web scraping, and a distilled RoBERTa model was fine-tuned to classify financial headlines into emotional categories. The fine-tuned model categorized headlines into emotion classes and assigned an emotion intensity scores. These extracted features were utilized as inputs for machine learning classifiers, including logistic regression, random forest, and artificial neural networks, to predict stock price movements. The researchers conducted several experiments to compare two approaches: one utilizing only emotion-based features from headlines, and another employing only traditional financial indicators such as price and volume data. They applied three machine learning classifiers (Logistic Regression, Random Forest, and Artificial Neural Network) to predict the next-day stock price direction. The results demonstrated that emotion-based features alone could achieve prediction accuracy comparable to financial indicator-based models. 

Haowei Ni et al. \cite{Ni2024} introduces an advanced LLM-based approach to predicting stock movements following earnings reports. Stock market movements are predicted using a comprehensive dataset that combines 'base factors' (financial metric growth and earnings transcripts) with 'external factors' (market indices and analyst grades). They collected data from 501 s\&p 500 companies, including financial reports, analyst sentiments, and market performance. This dataset was textualized, tokenized, and optimized for LLM processing. Various pre-trained models were fine-tuned using instruction-based training and QLoRA compression to enhance computational efficiency. The models were evaluated using accuracy, weighted F1 score, and Matthews correlation coefficient (MCC). The results showed that fine-tuned models, particularly llama-3-8b-Instruct-4bit, outperformed baseline models, such as GPT-4, achieving superior next-day stock movement predictions. 

In \cite{Li2024}, the Stock-Chain framework was introduced as a novel approach to enhancing financial analysis by integrating RAG with LLMs for stock trend prediction and financial question answering. To address these limitations, the authors present AlphaFin, a dataset that combines traditional financial research datasets, real-time market data, and manual CoT reasoning examples to improve LLMs' analytical abilities. The Stock-Chain framework leverages this dataset by fine-tuning an LLM-based model, StockGPT, which employs LoRA for efficient training. The framework integrates financial reports, market data, and financial news into a vector database, using retrieval-augmented techniques to enhance real-time knowledge retrieval and reasoning in financial predictions and question answering. Empirical evaluations of stock trend prediction and financial QA demonstrate that Stock-Chain outperforms classical ML models, general-purpose LLMs, and financial domain-specific LLMs (FinLLMs) in both accuracy and profitability. For stock trend forecasting, the model achieves an annualized return (ARR) of 30.8\% and an accuracy of 55.63\%, outperforming baselines through CoT-enhanced financial reasoning. In financial QA, Stock-Chain delivers superior results beacuse of its retrieval-augmented architecture, surpassing models such as FinGPT and ChatGPT in generating accurate and context-aware responses. Human and GPT-4 evaluations confirm its effectiveness, with win rates exceeding 60\% compared with other financial LLMs. 

Giorgos Iacovides  et al. \cite{Iacovides2024} introduces FinLlama, a financial sentiment analysis model that LLMs for enhanced algorithmic trading. It is based on Llama 2 7B, fine-tuned on 34,180 labeled financial text samples from four domain-specific datasets to classify sentiment polarity and quantify sentiment strength. The model employs a generator-discriminator architecture and LoRA to reduce training complexity while maintaining high performance. The framework includes a full pipeline comprising data collection, named entity recognition, text preprocessing, sentiment classification, and portfolio construction. A three-class softmax layer enables the model to distinguish between positive, neutral, and negative sentiments. LoRA-based fine-tuning allows FinLlama to retain efficiency while adapting effectively to finance-specific tasks. The performance of FinLlama was evaluated using real-world trading metrics, with a 35\% long-short portfolio constructed using FinLlama delivering higher cumulative returns, improved Sharpe ratio, and lower volatility than competing models such as FinBERT and lexicon-based methods.

A novel LLM-based framework \cite{Wang2024b}, StockTime, was introduced to enhance stock price predictions using LLMs for financial time-series forecasting. Stock prices are first segmented into patches and textual representations are extracted, encompassing correlations, trend movements, and timestamps. Temporal information from an autoregressive encoder was merged with these representations and integrated into the latent space of the LLM. The LLM was not fine-tuned in this approach; instead, only the embedding and projection layers were trained. The framework consists of modular components, including patched input segmentation, an autoregressive encoder, multimodal fusion, and token-level prediction, enabling StockTime to process structured time-series data while effectively minimizing the computational overhead. An empirical evaluation was conducted using six public financial datasets—s\&p 100-D, s\&p 100-H, Bigdata23, Bigdata22, ACL18, and CIKM18—spanning various timeframes and market conditions. The results show that StockTime consistently outperforms traditional time-series models and existing FinLLMs, delivering higher accuracy, lower computational costs, and better generalization. 

Kirtac and Germano \cite{Kirtac2025}proposed Sentiment-Augmented Proximal Policy Optimization (SAPPO), a novel framework that enhances traditional reinforcement learning-based portfolio optimization by integrating LLM-driven financial sentiment analysis. SAPPO enhances an agent’s observation space by using sentiment signals derived from financial news. This sentiment is extracted using a fine-tuned version of LLaMA 3.3, specialized for financial text, allowing the model to incorporate market-relevant qualitative insights directly into trading policy. The sentiment pipeline first retrieves daily financial news from Refinitiv, applies cosine similarity filtering to avoid redundancy, and then processes the selected articles using LLM to generate structured sentiment scores. These scores are appended to the state vector of the RL agent and used to adjust the policy's advantage function, enabling the agent to adapt dynamically to changing market sentiments. The method was evaluated using a three-stock portfolio (Google, Microsoft, and Meta) over the backtesting period from 2013 to 2020. Compared to standard PPO and financial benchmarks, SAPPO achieves a higher annualized return (30.2\%), sharper risk-adjusted performance (Sharpe ratio of 1.90), and reduced drawdown, indicating improved robustness in volatile markets. 

\begin{table}[!ht]
\caption{Agent-Based Architectures: Models, and Data, Applications}
\begin{center}
\begin{tabular}{|p{0.23\textwidth}|p{0.25\textwidth}|p{0.32\textwidth}|p{0.31\textwidth}|}
\hline
\textbf{Method/Paper} & \textbf{LLM Model} & \textbf{Data} & \textbf{Application} \\
\hline
Optimized AI-Agent collaboration \cite{Han2024} & GPT-4 & Dow Jones Index & Fundamental analysis, market sentiment analysis, and risk analysis \\
\hline
Alpha-GPT 2.0 \cite{Yuan2024} & Alpha-GPT & Historical market data, human interaction feedback, and modular AI components & Quantitative trading, portfolio management, and risk assessment \\
\hline
FS-ReasoningAgent \cite{Wang2024a} & GPT-3.5-turbo, GPT-4, GPT-4o, o1-mini & Cryptocurrency trading, Bitcoin (BTC), Ethereum (ETH), and Solana (SOL) & Cryptocurrency trading \\
\hline
FINCON \cite{Yu2025} & GPT-4-Turbo & Multi-modal dataset (2022--2023) & \begin{minipage}{0.3\textwidth}Stock trading and portfolio management\end{minipage} \\
\hline
\cite{Xiao2024} & gpt-40-mini and gpt-4o & Historical stock prices, news articles, social media posts, and earnings reports & Financial Trading \\
\hline
\raggedright Novel LLM-based Framework \cite{Kou2024} & Alpha Grail & Chinese stock market (SSE 50 Index) & Quantitative Stock Investment and Portfolio Management \\
\hline
MarketSenseAI2 \cite{Fatouros2025} & GPT-4o & S\&P 100 and S\&P 500 & Stock analysis \\
\hline
StockAgent \cite{Zhang2024b} & \raggedright GPT-3.5-Turbo, Gemini-Pro & Historical stock prices, financial reports, macroeconomic indicators, policy changes, market sentiment data, and transaction logs & Stock market trading \\
\hline
TwinMarket \cite{Yang2025} & GPT-4o & Real transaction records (Xueqiu), Guba Forum posts, Sina \& CNINFO news, SSE 50 stock data & Investor behavior simulation in a stock market environment \\
\hline
\cite{Henning2025} & Claude 3.5 Sonnet, GPT-4o, Grok-2, Mistral Large 24.11, GPT-3.5, and Gemini 1.5 Pro & - & Stock portfolio Investment \\
\hline
\cite{Ma2025} & \raggedright GPT-4o, Gemini-1.5, LLaMa-3, DeepSeek, Qwen-2.5 & NASDAQ & Stock portfolio Investment \\
\hline
\end{tabular}
\end{center}
\end{table}
\subsection{Agent-Based Architectures}
In \cite{Han2024}, a structured approach for designing AI agents was introduced that performs autonomously or collaboratively, enabling them to handle complex tasks, such as financial report analysis, through tool integration and multiagent coordination. Agents leverage the Text2Param capability of the underlying language model to dynamically invoke external tools, such as retrieving stock data or forum posts, and integrating RAG to autonomously process large documents without manual query adjustments. The framework supports different agent architectures, including single-agent systems that perform independently with function-calling capabilities and dual-agent collaboration, where enforced communication ensures mutual validation before finalizing outputs, and multi-agent structures that scale into horizontal, vertical, or hybrid collaboration models. In horizontal collaboration, agents engage in real-time discussions, while vertical collaboration introduces a leader-agent overseeing specialized subordinates. Hybrid collaboration combines these approaches, balancing shared decision-making with hierarchical oversight. They conducted experiments on SEC 10-K reports from 30 Dow Jones Index companies and demonstrated that single-agent models are highly effective in fundamental analysis, whereas multiagent structures yield superior performance in terms of risk assessment. The ensemble approach, which optimally selects agent structures based on task requirements, achieves an accuracy of 66.7\% in investment decision making and an average deviation of 2.35\% in stock price prediction, outperforming all other configurations.

Hang Yuan et al. \cite{Yuan2024} introduced Alpha-GPT 2.0, a Human-in-the-Loop AI framework for enhancing quantitative investment strategies by integrating human expertise with AI-driven financial modeling. Addressing the challenges of fully automated trading systems enables analysts to iteratively refine predictions and trading strategies through dynamic human-AI collaboration, improving adaptability, transparency, and accuracy in investment decision-making. The framework has three specialized agents: an alpha mining agent for extracting trading signals and patterns, an alpha modeling agent for developing predictive models and portfolio optimization, and an alpha analysis agent for risk assessment and fundamental analysis. Each agent uses LLMs and has specific tools, knowledge bases, and APIs. Alpha-GPT 2.0 relies on historical market data, human interaction feedback, and modular AI components to generate and refine investment strategies. The alpha mining agent maintains an alpha base with detailed annotations, the modeling agent maintains a collection of ML models and their configurations, and the analysis agent uses a financial behavior knowledge graph for risk assessment. User feedback dynamically guides the research process, while predefined tools for portfolio optimization and risk assessment act as reusable components for strategy refinement. The framework is applied to quantitative trading, portfolio management, and risk assessment, enabling investment firms to leverage AI for data-driven decision making.

A novel multiagent framework, Fact-Subjectivity-ReasoningAgent(FS-ReasoningAgent), was developed \cite{Wang2024a} to enhance LLM-based cryptocurrency trading by separately analyzing factual and subjective reasoning before making trading decisions. First, data were collected from two sources: market statistics (price, volume, and gas fees) from CoinMarketCap and cryptocurrency news from financial sources via the Gnews API.  The collected data were then processed by several specialized agents: a Statistics Agent for analyzing market metrics, a Fact Agent for processing objective news, a Subjectivity Agent for sentiment analysis, and dedicated reasoning agents for both factual and subjective components. After data processing, the reasoning step involved two agents. The Fact Reasoning Agent interprets factual data to provide logical trading insights, while the Subjectivity Reasoning Agent analyzes market sentiment and emotional trends. Their insights are then passed to the Trade Agent, which integrates both factual and subjective perspectives to determine whether to buy, sell, or hold a cryptocurrency. This decision-making process dynamically adjusts the weighting of factual and subjective data based on market conditions.  The proposed method was tested on Bitcoin (BTC), Ethereum (ETH), and Solana (SOL) trading data from November 2023 to July 2024 under both bull and bear market conditions. The results showed that FS-ReasoningAgent outperformed CryptoTrade and traditional trading strategies, demonstrating that subjective reasoning was more effective in bull markets, while factual data led to better results in bear markets.

In \cite{Yu2025}, FINCON was introduced as an LLM-based multi-agent framework designed for high-stakes financial decision making, specifically in single-stock trading and portfolio management. FINCON is designed to combine hierarchical agent organization, structured risk management, and adaptive investment belief updates to improve trading outcomes. FINCON integrates a Synthesized Manager-Analyst structure with a dual-level risk-control mechanism to optimize decision quality and efficiency. Analyst agents specialize in different market data modalities—textual financial reports, news sentiment, earnings call audio, and quantitative financial metrics—providing modular and denoised investment insights. The manager agent consolidates these insights, executes trading decisions, and continuously refines its investment beliefs through Conceptual Verbal Reinforcement (CVRF), which adapts its strategy based on past profit-and-loss (PnL) trends. To further enhance decision robustness, FINCON employs a within-episode risk control mechanism that monitors market fluctuations in real time using Conditional Value at Risk (CVaR) and an over-episode belief update mechanism that refines trading strategies episodically through text-based gradient descent. This combination enables FINCON to make adaptive risk-aware trading decisions while reducing over-reliance on static investment heuristics. The empirical evaluation demonstrates that FINCON outperforms both LLM-based and RL-based financial agent systems across multiple trading tasks. The framework was  trained and tested on a multimodal dataset (2022–2023) covering stocks TSLA, AMZN, NIO, MSFT, AAPL, GOOG, NFLX, and COIN, incorporating stock prices, financial news, earnings call recordings, and company filings. FINCON-based strategies outperform equal-weighted ETFs, state-of-the-art DRL agents and LLM-based trading models, mean-variance model and RL-based FinRL agents, and classical portfolio optimization techniques by optimizing asset allocation dynamically.

Yijia Xiao, et al. \cite{Xiao2024} propose TradingAgents, a stock trading framework that leverages specialized LLM agents such as fundamental analysts, sentiment analysts, technical analysts, and traders with varying risk profiles. It employs models such as gpt-40-mini and gpt-4o for low-depth tasks such as summarization, data retrieval, and converting tabular data to text, while o1-preview handles reasoning-intensive tasks like decision-making, evidence-based report writing, and data analysis. TradingAgents define seven distinct agent roles within a simulated trading firm: Fundamentals Analyst, Sentiment Analyst, News Analyst, Technical Analyst, Researcher Trader, Risk Manager, and Fund Manager. Each agent is assigned a specific role, goal, set of constraints, skills, and tools, enabling collaborative decision-making. Analysts gather and analyze market data from various sources, Researchers critically evaluate insights through structured debates, Traders make informed decisions based on research analysis, the Risk Manager assesses trading decisions against market conditions and the firm's risk tolerance, and finally, the Fund Manager reviews and approves the Risk Manager’s recommendations for execution. The framework integrates structured outputs with natural language reasoning for control and decision making. The dataset includes historical stock prices, news articles (from Bloomberg, Yahoo, FinnHub, Reddit), social media posts (X/Twitter, Reddit), and earnings reports. 

Kou et al., \cite{Kou2024} introduces a framework that combines Large Language Models and a multiagent system to improve quantitative stock investment strategies. The methodology begins with the extraction and generation of alpha factors from diverse financial data sources including numerical data, research papers, and visual charts. Alpha Grail" (a custom version of ChatGPT) is utilized to process and categorize these factors according to established financial categories. After the generation of alpha factors, a multi-agent system evaluates its effectiveness under different market conditions. Agents perform the following tasks as part of the multi-agent system: they analyze historical data to evaluate the effectiveness of alpha factors and identify those that demonstrate strong predictive power. They filtered out underperforming factors to ensure that only the most reliable ones were retained. They assigned confidence scores to the selected alpha factors based on their predictive accuracy, prioritizing those most relevant to current market conditions. Agents continuously monitor financial environments and adjust their evaluations to dynamically refine their investment strategies. This structure enables the system to adjust strategies dynamically as market conditions shift. After selecting the most significant alpha factors, the framework utilizes a deep learning model to determine the optimal weights for each factor. This model predicts the future performance of individual alpha factors and assigns them appropriate weights to create a flexible investment strategy. The proposed framework was tested on stock market data from the Chinese stock market (SSE 50 Index, January 1, 2021, to December 31, 2023) and outperformed existing quantitative models in various financial metrics. 

MarketSenseAI 2.0 introduced in \cite{Fatouros2025} is a holistic stock analysis framework leveraging Large Language Models (LLMs) to integrate financial news, historical prices, company fundamentals, and macroeconomic data to improve stock selection and predictive accuracy. It is designed for investment signal generation by processing diverse financial inputs through specialized AI agents. The framework introduces a CoA approach for granular fundamental analysis and a RAG module using Hypothetical Dense Embeddings (HyDE) for enhanced macroeconomic insights. The system employs five AI agents: News Agent, Fundamentals Agent, Dynamics Agent, Macroeconomic Agent, and Signal Agent, each analyzing a specific aspect of financial data. These agents extract, process, and interpret real-time financial signals to generate actionable stock recommendations. It uses GPT-4o as the primary LLM, VectorBTPro for portfolio analysis, and multiple data-processing tools. Data sources include s\&p 100 (2023-2024) and s\&p 500 (2024) stock data, incorporating historical stock prices, financial reports, market indices, and macroeconomic indicators to refine decision making. Performance evaluation is conducted using cumulative returns, alpha generation, and factor analysis, with results showing 125.9\% cumulative returns in s\&p 100, outperforming the s\&p 100 index return of 73.5\%.

Chong Zhang et al. \cite{Zhang2024b} introduces StockAgent, a multi-agent AI system that simulates stock trading using Large Language Models (LLMs) like GPT-3.5-Turbo and Gemini-Pro. It examines how AI agents replicate real-world trading environments and evaluates their responses to macroeconomic factors, market trends, and investor sentiment. This research focuses on three key areas: simulation effectiveness, LLM reliability in trading decisions, and the impact of external conditions on AI-driven stock trading. StockAgent comprises three core modules: the Investment Agent module, which assigns distinct risk profiles to AI agents; the Transaction Agent module, which processes buy/sell decisions and updates market prices; and the Bulletin Board System (BBS) module, where agents exchange insights and simulate market sentiment. The system simulates 264 trading days and incorporates historical stock price data, financial reports, macroeconomic indicators, and policy changes to create a dynamic market environment. Performance is evaluated using stock price correlation, transaction volume, profitability metrics, and statistical analysis, utilizing data from historical stock price movements, financial reports (balance sheets, income statements, cash flow statements), macroeconomic indicators (interest rate changes, monetary policy shifts), market sentiment data (BBS discussions), and transaction logs. The findings reveal that GPT-driven agents adopt aggressive strategies with higher volatility, while Gemini-based agents trade more conservatively. Additionally, removing key data sources, such as interest rates or BBS discussions, significantly alters trading behavior.

Congluo Xu et al. [34] introduced FinArena, a human-agent collaboration framework developed to enhance financial market analysis and forecasting through multimodal data integration and personalized recommendations. Inspired by the MoE architecture, it combines stock prices, news articles, and financial statements within an LLM-based multiagent system. To mitigate LLM hallucinations, the framework incorporates an adaptive retrieval-RAG method to process unstructured news content and enrich it with relevant external information. FinArena aims to deliver tailored investment decisions by incorporating investors’ individual risk preferences. The system consists of three specialized agents and one central analytical unit. The time series agent forecasts stock prices using historical time-series data. The News Agent extracts insights from financial news via the adaptive RAG method, improving relevance and factual accuracy. The Statement Agent applies iterative reasoning to structured financial statements to extract key metrics. These outputs are synthesized by an AI expert, who aligns investment strategies with user-defined risk tolerance and produces personalized recommendations. This framework was evaluated on stock trend predictions and trading simulations across multiple investor profiles. The results show FinArena outperforms traditional and state-of-the-art models, especially in the U.S. market, and generates more accurate forecasts and investment guidance. FinArena highlighted the importance of accessible AI-driven tools for retail investors and the value of aligning strategies with individual risk preferences. 

Yang et al. \cite{Yang2025} introduced TwinMarket, a comprehensive simulation framework that applies LLM-driven agent modeling to analyze how micro-level investor behaviors lead to macro-level market outcomes. TwinMarket builds upon the Belief-Desire-Intention (BDI) model to simulate autonomous agents that perceive market conditions, reason through beliefs and intentions, and perform trading actions accordingly. Each agent uses GPT-4o to simulate core investor behaviors such as overconfidence, loss aversion, anchoring bias, herding, and rationality decay. TwinMarket has a simulation environment comprising a dynamic financial world with a stock market, news system, and social interaction network. Agents engage in multi-stage decision-making, influenced by real-time company fundamentals (from CSMAR), financial news (from Sina and CNINFO), and social signals (from the Guba forum posts). To ground agent behavior, the framework uses real transaction data from Xueqiu to construct initial user profiles based on actual investment patterns, demographics, and risk preferences. These agents then interact through a social network, forming and adjusting beliefs based on the rumor spread, news sentiment, and peer influence. The system simulates trading activity across ten indices derived from the SSE 50 constituents and scales from 100 to 1,000 agents, maintaining robust performance. TwinMarket can reproduce a variety of real-world financial dynamics, including fat-tailed return distributions, volatility clustering, boom-bust cycles, self-fulfilling prophecy effects, and information-driven market crashes. The framework also demonstrates how social dynamics, such as the spread of rumors or trust in dominant voices, can shift market sentiment and lead to irrational collective behavior.  

Thomas Henning et al. \cite{Henning2025}[48] explore how LLMs behave in a classic experimental finance paradigm designed to understand how and why such speculative bubbles and crashes phenomena emerge. The study investigates LLM-based agents in in two market settings: single-model markets where all traders are instances of the same LLM and mixed-model "battle royale" settings, where multiple LLMs compete within the same market. In single-model markets, three separate experiments were conducted using agents based on six commercial LLMs: Claude 3.5 Sonnet, GPT-4o, Grok-2, Mistral Large 24.11, GPT-3.5, and Gemini 1.5 Pro. Each market consisted of 20 identical instances of a given LLM, with agents receiving the same prompt and acting independently to maximize their profit, unaware of their counterparties’ identities. The experiment determines whether the market followed rational investment principles keeping asset prices aligned with fundamental value or if it exhibited human-like behaviors such as speculation, trend-following, and bubble formation. In the mixed-model setup, the six LLMs competed directly in a 24-agent market with four agents per model to evaluate their relative performance. The performance evaluation relied on the Mean Squared Error (MSE) and Pearson Correlation Coefficient (PCC). The findings indicate that LLMs generally exhibit a "textbook-rational" approach, keeping prices close to fundamental value and showing only a limited tendency toward bubble formation. The results indicate that some models, such as Claude-3.5-Sonnet and GPT-4o, exhibit more rational forecasting behavior than others such as GPT-3.5. 

Tianmi Ma, et. al. \cite{Ma2025} introduce the Agent Trading Arena, a dynamic, zero-sum trading environment designed to evaluate LLMs' numerical reasoning beyond predefined problem-solving. LLM-based agents perform stock analysis, decision-making, and reflection, adapting strategies through trial and error. Each agent's workflow integrates LLMs for chat pool interactions, stock analysis, decision-making, and reflection. The agent responsible for generating actions receives prompts and decides whether to buy, sell, or hold stocks based on input information. To enhance the performance of LLMs to capture market trends and patterns, this study shifted from textual numerical data to visual representations, such as scatter plots and graphs. Additionally, a reflection module enhances learning by analyzing performance, providing real-time feedback, and iteratively refining strategies. This study evaluates models such as GPT-4o, Gemini-1.5, LLaMa-3, DeepSeek and Qwen-2.5. Using historical NASDAQ data for validation, the findings demonstrate that visual data and the reflection process significantly improve LLMs' numerical reasoning, allowing better data analysis and informed trading decisions. Results indicate that visual data and the reflection process significantly enhance LLMs' numerical reasoning, leading to improved data analysis and trading decisions. 
\section{Conclusions and future directions}
The integration of LLMs into financial investment strategies has shown significant promise in enhancing decision making, improving predictive accuracy, and generating actionable insights. This paper reviews recent approaches across four major frameworks: LLM-based frameworks and pipelines, hybrid integration methods, fine-tuning and adaptation approaches, and agent-based architectures. Each method demonstrates unique strengths in leveraging LLMs for financial tasks such as risk assessment, portfolio optimization, and stock prediction. While the integration of LLMs into financial investment strategies has demonstrated remarkable progress, several areas remain open for further exploration and development. Future research could focus on the following directions. 
\begin{itemize}
    \item \textbf{Adaptive Fine-Tuning and Continuous Learning:} While current fine-tuned models such as GPT-InvestAR and StockGPT have shown strong predictive performance, they often face challenges adapting to evolving financial conditions. Future work could explore reinforcement learning, online fine-tuning strategies, and adaptive prompt engineering to ensure models remain relevant in dynamic market environments. 
    \item \textbf{Specialized LLM Architectures for Financial Analysis:} While general-purpose LLMs such as GPT-4 have been successfully adapted for financial tasks, the development of domain-specific LLM architectures optimized for financial reasoning is a key area for further research. Future models should incorporate: 1) Temporal modeling techniques tailored to financial time series data. 2) Enhanced data fusion methods to integrate numerical, textual, and visual financial information. 3) Domain-adaptive pre-training strategies that leverage financial language corpora to improve contextual understanding. 
    \item \textbf{Enhanced Multi-Agent Financial Systems:} Inspired by systems such as FINCON and StockAgent, future research should investigate more sophisticated agent architectures that enable adaptive specialization of agents for distinct financial tasks, dynamic communication frameworks for collaborative analysis, and hierarchical decision-making protocols that mimic the structure of expert financial teams. 
    \item \textbf{Adaptive Learning for Dynamic Markets:}The rapid evolution of financial markets demands LLM architectures that can be adapted in real time. Future work could explore self-supervised learning mechanisms for continuous adaptation, online learning frameworks that dynamically adjust prediction models, and memory-augmented architectures that efficiently combine historical and emerging data.
    \item \textbf{Comprehensive Benchmarking Frameworks:}To improve comparability among financial LLM models, future research should focus on developing standardized benchmark datasets tailored to financial prediction tasks and establishing evaluation metrics that capture both predictive accuracy and investment outcomes. 
    \item \textbf{Human-AI Collaboration in Financial Decision Making:}Building on Alpha-GPT 2.0 and similar frameworks, future research could explore improved methods for human-AI collaboration, including: 1) Enhancing user interfaces that facilitate efficient communication between analysts and LLM systems. 2) Incorporating expert feedback into financial models. 3) Developing personalized financial assistance systems tailored to individual investor profiles.
\end{itemize}

\vspace{2cm}

\end{document}